\documentclass[10pt,conference]{IEEEtran}
\usepackage{threeparttable}
\usepackage{multirow}
\usepackage{booktabs}
\usepackage{graphicx}
\usepackage{xspace}
\usepackage{dsfont}
\usepackage{algorithm}
\usepackage[noend]{algpseudocode}
\usepackage[normalem]{ulem}
\useunder{\uline}{\ul}{}
\usepackage{ragged2e}
\usepackage{tcolorbox}
\usepackage{array}
\usepackage{amsmath}
\usepackage{balance}
\usepackage{booktabs}
\usepackage{caption}
\usepackage{hyperref}
\usepackage{threeparttable}
\usepackage{enumitem}
\usepackage{pifont}
\usepackage{bbding}
\usepackage{xfrac}
\usepackage{diagbox}
\usepackage{amssymb}
\usepackage{graphicx}
\usepackage{subcaption}
\usepackage{cite}

\algrenewcommand\textproc{\texttt}

\usepackage[compact]{titlesec}         
\titlespacing{\section}{5pt}{6pt}{5pt} 
\AtBeginDocument{%
  \setlength\abovedisplayskip{0pt}
  \setlength\belowdisplayskip{0pt}}
\algrenewcommand\textproc{\texttt}

\renewcommand{\Comment}[1]{}

\newcolumntype{C}[1]{>{\centering\arraybackslash}p{#1}}
\usepackage{tikz}
\definecolor{customblue}{RGB}{192, 208, 235}
\definecolor{customdarkblue}{RGB}{68, 114, 196}

\AtBeginDocument{%
  \providecommand\BibTeX{{%
    \normalfont B\kern-0.5em{\scshape i\kern-0.25em b}\kern-0.8em\TeX}}}

\begin{document}

\title{Beyond Confidence: Rethinking Uncertainty Calibration in Deep Code Models}

\author{Anonymous Authors}



\makeatletter 
\newcommand{\linebreakand}{%
  \end{@IEEEauthorhalign}
  \hfill\mbox{}\par
  \mbox{}\hfill\begin{@IEEEauthorhalign}
}
\makeatother 

\author{
\IEEEauthorblockN{Ravishka Rathnasuriya}
\IEEEauthorblockA{Department of Computer Science \\ University of Texas at Dallas \\
ravishka.rathnasuriya@utdallas.edu}

\and

\IEEEauthorblockN{Wei Yang}
\IEEEauthorblockA{ Institute of Systems for Advanced Computing\\ Fudan University 
\\
yang\_wei@fudan.edu.cn}
}

\maketitle
\begin{abstract}

Deep learning models have shown significant promise in code-related tasks. While confidence calibration techniques have been proposed to improve model reliability, their effectiveness in real-world tasks with strict accuracy requirements remains uncertain. In this paper, we investigate two key questions: first, whether existing measurement metrics, including confidence scores and uncertainty metrics, are effective for guiding selective prediction; and second, whether existing calibration techniques can improve these metrics to better identify cases
where model predictions are likely to succeed or fail.

We conduct a systematic study on defect prediction and vulnerability detection using seven code models, evaluating thirteen uncertainty signals and three widely used post-hoc calibration methods. Our analysis reveals three major findings: (1) even strong uncertainty signals can remain poorly calibrated and provide limited separation at the high-confidence end, constraining selective prediction in practice; (2) standard post-hoc calibrators often improve calibration scores but do not consistently translate into better selective prediction behavior; and (3) commonly used calibration metrics can disagree with selective prediction performance, underscoring the need to evaluate both jointly. Moreover, our findings motivate a decision-oriented view of reliability in deep code models, where calibrated scores must be evaluated together with selective-prediction quality and retained-region risk before they can support accept--defer deployment.

\end{abstract}
\section{Introduction}\label{sec::intro}

Deep learning code models have demonstrated impressive capabilities in various code-related tasks such as defect prediction and vulnerability detection~\cite{tian2023fly, naturalattack, Zhang2023Challenging, lu2021codexglue, yefet2020adversarial,hu2023codes,li2021estimating,van2020tailoring,peng2018t}. To improve the reliability of these models, recent research
has explored confidence calibration techniques that align predicted probabilities with true correctness likelihoods~\cite{vasilev2023calibration,gupta2020calibration,krishnan2020improving,spiess2024calibration,zhou2024calibration}.
However, the practical effectiveness of these techniques in code tasks remains unclear, particularly when models are used in automated systems where inaccurate predictions can lead
to costly failures. In such cases, low-confidence predictions are often deferred to human experts. This paper investigates whether existing calibration methods genuinely enhance code model reliability or whether alternative approaches provide more trustworthy guidance.

While recent techniques~\cite{zhou2024calibration,spiess2024calibration,virk2024enhancing} have been proposed to improve the confidence calibration on code models, these techniques often treat calibration accuracy as an end in itself rather than as a tool to enhance downstream performance. These techniques primarily aim to align predicted confidence scores with empirical correctness, yet these studies seldom evaluate whether better-calibrated confidence actually improves the utility of code models in practice. In particular, these studies overlook whether calibration improvements lead to gains in accuracy or selective prediction~\cite{xin2021art, calder1999selective}, where the model produces an output only when it is sufficiently confident, and otherwise abstains to avoid unreliable predictions. In contrast, our work extends beyond calibration by jointly examining selective prediction and reliability assessment, two concerns that are essential for reliable deployment but often neglected in prior studies. This broader focus helps clarify how calibration can support trustworthy and actionable predictions in real-world settings.

Our investigation is motivated by two key concerns. First, we examine whether existing confidence and uncertainty scores are reliable indicators of prediction quality in code models. Unlike models in other domains, code models face unique challenges that affect their trustworthiness. One such issue is their tendency to rely on superficial patterns rather than truly understanding the semantics of the code, which can lead to misleading predictions. These models may confidently label a vulnerable function as safe simply because it resembles patterns seen in non-vulnerable code. This becomes especially problematic when the vulnerability is hidden within well-structured code that appears correct on the surface. Another challenge is that models often focus on local syntax but fail to capture important relationships across different parts of the codebase, which further reduces their ability to make reliable predictions. In such cases, relying only on confidence or uncertainty scores may not reflect when the model is making a mistake. Even if the score is well-calibrated, it might still assign high confidence to an incorrect prediction. This raises the question of whether existing~\cite{hendrycks2018baseline,steinhardt2016unsupervised,shannon1948mathematical,monarch2021human,gal2016dropout,lakshminarayanan2017simple,CodeImproveicse,rathnasuriya2026answer,rathnasuriya2026fly} or new metrics tailored for code model behavior could offer better signals for deciding when model outputs can be trusted.

Even if confidence or uncertainty scores are accepted as reasonable inputs for calibration, widely used calibration metrics face important limitations. Metrics like Expected Calibration Error (ECE)\cite{guo2017calibration} and Uncertainty Calibration Error (UCE) \cite{laves1909well} focus on aligning predicted probabilities with true correctness across the entire confidence range. However, this broad averaging places equal weight on all predictions, including low-confidence ones that are often ignored in real applications. In practice, such predictions are typically treated as unreliable by default, regardless of how well they are calibrated. What matters more is the accuracy of high-confidence predictions, since these are the outputs that users are most likely to trust. For code-related tasks, errors in this high-confidence region are especially costly. Yet, current calibration metrics often under-represent these cases, providing limited insight into how well the model performs when it is most certain. As a result, improvements in overall calibration may obscure critical failures in the predictions that matter most for reliable deployment.

To address the limitations of existing work, this paper investigates the practical reliability of calibrated deep code models through a focused study of \emph{code classification} tasks. We intentionally restrict our empirical evaluation to defect prediction and vulnerability detection because they provide a clean setting in which correctness and uncertainty can be defined at the example level. This allows us to assess behavioral calibration and selective prediction without conflating the analysis with unresolved measurement issues in generative code tasks. At the same time, this setting remains important in its own right as code classification is a core form of \emph{code understanding}, and it provides a meaningful basis for evaluating how well code models distinguish reliable from unreliable predictions in software engineering workflows.

Within this scope, we study two related questions. First, we evaluate whether current uncertainty signals are effective for selective prediction, that is, whether they can distinguish which predictions a model should make and which it should defer (Section~\ref{sec::emperical}). Second, we examine whether standard post-hoc calibration techniques improve a model's ability to separate reliable from unreliable predictions. While these methods are designed to align predicted probabilities with empirical correctness, their value for deployment-oriented decision-making in code tasks remains unclear.

Through a systematic study across SE tasks, we assess how uncertainty estimation and calibration influence a model's capacity to support selective prediction. Our study yields three takeaways about reliability in deep code classification. First, uncertainty signals can remain poorly calibrated and often fail to reliably prioritize the highest-confidence predictions, which are most critical for practical deployment. This limits how effectively selective prediction can stratify outputs by reliability. As a result, selective prediction gains can plateau and become unstable at high rejection, where decisions are driven by a small retained subset and minor ranking errors have outsized impact. Second, post-hoc calibration improves probability alignment but does not guarantee improved selective decision quality. Even when calibration metrics improve, models can continue to produce high-confidence mistakes, which directly undermines accept--defer behavior. Third, commonly used evaluation metrics provide incomplete guidance when used in isolation. Global calibration scores summarize average alignment but can miss the failure mode most relevant to deployment, namely whether errors are assigned low confidence. This motivates reporting calibration and selective prediction performance jointly, and analyzing how ranking quality changes across the confidence spectrum.


Building on these empirical results, the paper develops a broader account of reliability in deep code models. We show that reliability failure is shaped jointly by data ambiguity, representational bias, optimization objectives, and deployment-time decisions, rather than by score miscalibration alone. This explains why better global calibration does not always translate into safer selective prediction. More importantly, it motivates a shift in software engineering research toward decision-oriented reliability methods and evaluation criteria that emphasize accepted-set safety and behavior in the high-confidence region.

In summary, this paper makes the following contributions:
\begin{itemize}
    \item \textbf{Joint focus on calibration and abstention: } We study uncertainty calibration together with selective prediction, which are ignored in prior calibration work.
    \item \textbf{Empirical study:} A large-scale evaluation of 13 uncertainty metrics and three post-hoc calibration methods across seven code models and two SE tasks, revealing persistent miscalibration, especially in high-confidence regions.
    \item \textbf{Analysis of metric limitations:} We show that common calibration metrics (e.g., ECE) often fail to reflect true prediction reliability. 
    
    
    \item \textbf{Operationalizing uncertainty for SE:} We show how uncertainty should be evaluated for accept--defer deployment using calibration quality, selective-prediction quality, and retained-region risk.
    





\end{itemize}


\section{Preliminaries}
\label{sec::background}

Deep learning models are increasingly used for software engineering classification tasks such as vulnerability detection and defect prediction~\cite{tian2023fly, naturalattack, Zhang2023Challenging, lu2021codexglue, yefet2020adversarial, hu2023codes, li2021estimating, van2020tailoring, peng2018t}. Despite strong predictive performance, these models can produce overconfident errors~\cite{zhou2024calibration, spiess2024calibration}, assigning high confidence to incorrect labels. Such behavior is problematic in automated workflows, where predictions may trigger downstream actions in security and quality assurance settings.


Reliability in these settings requires more than accuracy. It also requires confidence or uncertainty signals that reflect the likelihood of correctness, so that a system can decide when to trust a prediction and when to defer it. Selective prediction formalizes this requirement by allowing a model to abstain on inputs where its decision is unreliable~\cite{xin2021art, calder1999selective}. A common implementation accepts predictions whose confidence exceeds a threshold and defers the rest, trading coverage for accuracy on the accepted subset. The effectiveness of this approach depends on whether the underlying confidence or uncertainty signal meaningfully separates correct from incorrect predictions.

\subsection{Calibration}

Calibration refers to the alignment between a model’s predicted confidence scores and the actual likelihood of correctness~\cite{guo2017calibration, zhou2024calibration, spiess2024calibration, vasilev2023calibration, gupta2020calibration}. For classification models that output a predictive distribution $P_M(\cdot \mid X)$, the predicted label is $\hat{y}=\arg\max_y P_M(y\mid X)$ and the associated confidence is $c(X)=\max_y P_M(y\mid X)$.
A model is perfectly calibrated if, for every confidence level $p \in [0,1]$, the predictions made with confidence $p$ are correct with frequency $p$:
\begin{equation}
\Pr(\hat{y} = y \mid c(X) = p) = p, \quad \forall p \in [0,1].
\end{equation}
Because exact conditioning on $c(X)=p$ is not attainable with finite samples, calibration is typically assessed using summary metrics. The Brier score measures the mean squared error between predicted probabilities and the observed labels, while ECE estimates the average gap between confidence and accuracy over confidence bins~\cite{guo2017calibration, gupta2020calibration}. These metrics are widely used to compare models and to evaluate post hoc calibration methods.

\subsection{Uncertainty Calibration for Selective Prediction in Code Tasks}

In code classification, confidence can be unreliable even when accuracy is high. Predictions may be driven by lexical or syntactic cues that correlate with labels in the training data but fail to reflect program semantics~\cite{Devlin2019BERT,Sanh2020DistilBERT,guo2021graphcodebert,Chen2021Evaluating}. As a result, a model can assign high confidence to a misclassification when the input contains subtle behaviors that are not captured by surface features. Long-range dependencies and interactions across statements further complicate this setting, since errors may depend on relationships that are difficult to represent with local token evidence.

These properties motivate studying uncertainty signals beyond the maximum softmax probability. Prior work has proposed entropy-based and sampling-based measures that aim to capture predictive ambiguity and model variability~\cite{hendrycks2018baseline,steinhardt2016unsupervised,shannon1948mathematical,gal2016dropout,lakshminarayanan2017simple,CodeImproveicse}. In our setting, these measures are used as scoring functions for selective prediction, where larger uncertainty is intended to indicate higher error likelihood. Uncertainty calibration then refers to aligning these scores with empirical correctness so that thresholds correspond to predictable reliability levels. Metrics such as ECE~\cite{guo2017calibration,gupta2020calibration,vasilev2023calibration} and UCE~\cite{laves1909well} are commonly used to quantify this alignment.

Poor calibration weakens selective prediction in two ways. First, overconfident errors may remain among accepted predictions, undermining reliability. Second, overly conservative scores may defer many correct predictions, reducing coverage without improving decision quality. Understanding when uncertainty signals are well calibrated for code models, and when calibration methods and metrics fail to reflect decision utility, is therefore necessary for reliability-oriented deployment.

\section{Research Methodology}
\label{sec::methodology}


This section details our experimental methodology for evaluating uncertainty signals and calibration.  We present the datasets and models studied, uncertainty estimation techniques, the post-hoc calibration methods, and evaluation metrics employed in this study.

\subsection{Subjects}
\subsubsection{Datasets and Tasks}

We study uncertainty and calibration in two widely used \emph{code classification} tasks, vulnerability detection~\cite{zhou2019devign} and defect prediction~\cite{phan2017conv}. We focus on classification for three reasons: (i) these tasks are widely used in practice, (ii) uncertainty is easy to obtain from predictive probabilities without additional inference, and (iii) confidence thresholds map directly to accept-or-defer decisions in selective prediction.

\textbf{Vulnerability detection (binary classification).} We use the Devign dataset~\cite{zhou2019devign}, which contains 27,318 C functions from FFmpeg and Qemu labeled as \emph{vulnerable} (14,858) or \emph{clean} (12,460). We use the standard split into training (21,854), validation (2,732), and test (2,732) sets.

\textbf{Defect prediction (multi-task classification).}
We use the CodeChef dataset~\cite{phan2017conv}, which contains 33,822 C/C++ functions labeled as \emph{no defect} (11,362), \emph{wrong output} (13,656), \emph{timeout error} (5,101), or \emph{runtime error} (3,703). We use the provided split into training (21,647), validation (5,411), and test (6,764) sets.





\subsection{Models}

We use seven state-of-the-art pre-trained models: CodeBERT~\cite{fengCodeBERTPreTrainedModel2020}, RoBERTa~\cite{Liu2019RoBERTa},  GraphCodeBERT~\cite{guo2021graphcodebert}, 
StarCoder-3B~\cite{li2023starcoder}, CodeLlama-7B~\cite{roziere2023code}, DeepSeek-Coder-7B~\cite{guo2024deepseek}, and Qwen2.5-Coder-7B~\cite{hui2024qwen2}. These models are fine-tuned on the respective tasks following standard practices~\cite{ding2024vulnerability,yin2024multitask}. Their performance in terms of accuracy (A), precision (P), recall (R), and F1-scores (F1) is summarized in Table~\ref{Tab:performance}. These models are widely used in prior work~\cite{lu2021codexglue,tian2023code,yang2022natural} and collectively cover diverse architectures, including encoder-only and decoder-based code LLMs, providing a representative view of modern code classification models.

\begin{table}[htb!]
\caption{Overall performance of code models.}
\label{Tab:performance}
\centering
\renewcommand{\arraystretch}{}
\resizebox{\linewidth}{!}{
\begin{tabular}{l c c c c c c c c}
\hline
\multirow{2}{*}{\textbf{Model}} 
& \multicolumn{4}{c}{\textbf{Vulnerability Detection}} 
& \multicolumn{4}{c}{\textbf{Defect Prediction}} \\
\cmidrule(lr){2-5} \cmidrule(lr){6-9} 
& \textbf{A} & \textbf{P} & \textbf{R} & \textbf{F1}
& \textbf{A} & \textbf{P} & \textbf{R} & \textbf{F1} \\
\hline
CodeBERT & 63.32 & 64.56 & 63.36 & 61.07 & 84.41 & 84.39 & 84.17 & 84.16 \\
 
RoBERTa &  61.96 & 61.84 & 61.96 & 61.18 & 82.17 & 81.96 & 82.03 & 81.90 \\
GraphCodeBERT & 64.31 & 64.76 & 64.31 & 63.04 & 85.4 & 85.34 & 85.40 & 85.20 \\
StarCoder-3B        & 59.69 & 59.01 & 40.15 & 47.79 & 76.24 & 75.82 & 76.24 & 75.19 \\
Qwen2.5-Coder-7B    & 61.53 & 59.27 & 51.95 & 55.37 & 77.89 & 77.34 & 77.91 & 77.48 \\
CodeLlama-7B        & 62.51 & 60.12 & 54.66 & 57.26 & 77.91 & 77.16 & 77.91 & 77.33 \\
DeepSeek-Coder-7B   & 59.59 & 57.51 & 46.05 & 51.15 & 79.77 & 79.29 & 79.77 & 79.47 \\
\hline
\end{tabular}}
\end{table}



\subsection{Uncertainty Measurement Metrics}


We analyze 13 uncertainty metrics covering aleatoric and epistemic uncertainty. Confidence-based metrics such as \textbf{vanilla}~\cite{hendrycks2018baseline} which uses the highest predicted probability as confidence; margin confidence \textbf{(MC)} and ratio confidence \textbf{(RC)}~\cite{monarch2021human} that measure the difference between the top two probabilities, with smaller gaps indicating higher uncertainty; and least confidence and Best-vs-Second-Best \textbf{(BvSB)}~\cite{monarch2021human} that assess uncertainty by evaluating class score differences. The metrics are widely used due to their simplicity, but they often produce misleading results in code tasks, where models may assign high confidence to semantically incorrect predictions driven by surface-level syntax.

Information theory-based metrics~\cite{steinhardt2016unsupervised,shannon1948mathematical} such \textbf{entropy} that measures the dispersion of the probability distribution and mutual information \textbf{(MI)} which captures the difference between total and expected entropy. These metrics may fail when softmax outputs are sharply peaked but misaligned with ground truth. To address these limitations, we also include bayesian approximation-based metrics such as Monte Carlo Dropout \textbf{(MCD)}~\cite{gal2016dropout} which estimates uncertainty by performing multiple stochastic forward passes and Deep Ensembles \textbf{(DE)}~\cite{lakshminarayanan2017simple} which aggregates predictions from multiple independently trained models. We derived three different measures from \textbf{MCD} and \textbf{DE} namely, (1) Sample winning score \textbf{(SWS)} which measures prediction consistency across multiple forward passes; (2) Probability variance \textbf{(PV)} which evaluates variance in softmax probabilities; and (3) Bayesian active learning by disagreement \textbf{(BALD)} which quantifies epistemic uncertainty. By comparing these metrics, we aim to understand which ones give reliable uncertainty estimates for code-related tasks, where models often behave unpredictably.

\subsection{Calibration Methods }


We employ three post-hoc calibration techniques: Temperature Scaling (TS) \cite{guo2017calibration, zhou2024calibration}, Platt Scaling (PS)\cite{spiess2024calibration}, and Isotonic Regression (Iso.R)~\cite{guo2017calibration}, which have been widely adopted in prior studies~\cite{guo2017calibration,zhou2024calibration,spiess2024calibration,gupta2020calibration,vasilev2023calibration,laves1909well}.




\textbf{TS}~\cite{guo2017calibration, zhou2024calibration} applies a single scaling parameter to smooth confidence scores, reducing overconfidence in predictions. 
Although simple and effective for correcting mild miscalibration in balanced datasets, TS assumes uniform error across samples and may fail when confidence varies significantly across subgroups or semantic classes. We adapt TS to rescale uncertainty scores via:
\begin{equation} \tilde{u}_i = \frac{u_i}{T} \end{equation}

where \( u_i \) is the original uncertainty score for sample \( i \), \( T \) is the learned temperature parameter, and \( \tilde{u}_i \) is the calibrated uncertainty score. Unlike conventional TS, which optimizes Negative Log-Likelihood, we minimize Binary Cross-Entropy (BCE) to suit binary correctness signals. TS is computationally efficient, but it often fails to correct miscalibration in unstable prediction settings, which are common in code classification tasks.

\textbf{PS} extends TS by applying a logistic transformation with two learnable parameters:
\begin{equation}
p_i = \sigma(a u_i + b)
\end{equation}
where ${u_i}$ is the original uncertainty score, 
$\sigma(x)$ is the sigmoid function, and ${p_i}$ is the calibrated probability. The $a$ and $b$ are learned parameters. To avoid overfitting, we use five-fold rescaling, iteratively training on 80\% of the data and validating on 20\%. However, PS assumes a globally smooth calibration curve and may extrapolate poorly on non-monotonic uncertainty patterns, which frequently arise in code tasks. 





\textbf{Iso.R}\cite{guo2017calibration} is a non-parametric calibration method that learns a monotonic function mapping uncertainty scores to empirical correctness. Unlike TS and PS, Iso.R does not assume a predefined functional form, making it highly flexible but prone to overfitting when data is sparse. 

The learned function $f(u)$ minimizes calibration error:
\begin{equation}
\hat{f} = \underset{f \text{ is isotonic}}{\arg \min} \sum_{i} (f(u_i) - y_i)^2
\end{equation}
where \( u_i \) represents the original uncertainty score, \( y_i \) is the observed correctness (binary label), and \( \hat{f} \) is the learned isotonic function constrained to be monotonic.

\noindent\textbf{Calibration fitting and evaluation.}
We follow a standard post-hoc calibration procedure to avoid test-set leakage~\cite{guo2017calibration,zhou2024calibration,spiess2024calibration}. For each subject model, the classifier is trained using the training split. We then fit each calibrator using held-out validation predictions and their corresponding correctness labels. After fitting, the calibration mapping is fixed and applied once to the test split. All reported calibration metrics and selective-prediction results are computed on the test split, which is never used to train, tune, or select calibration parameters.

\subsection{Calibration Metrics}


We evaluate calibration effectiveness using metrics that quantify the alignment between predicted uncertainty and actual correctness, assessing how well uncertainty estimates reflect model reliability. These metrics are widely used in prior studies to benchmark calibration performance across domains, offering standardized and interpretable measures of miscalibration~\cite{guo2017calibration,zhou2024calibration,spiess2024calibration,gupta2020calibration,vasilev2023calibration,laves1909well}.

\textbf{Expected Calibration Error (ECE)}~\cite{guo2017calibration} measures the average deviation between predicted confidence and empirical accuracy across \( M \) bins.
\begin{equation}
\mathit{ECE} = \sum_{m=1}^{M} \frac{|B_m|}{n} \left| \text{acc}(B_m) - \text{conf}(B_m) \right|
\end{equation}
where \( B_m \) is the set of samples whose confidence falls into the \( m^{th} \) bin,  \( |B_m| \) is the number of samples in bin \( m \),  \( n \) is the total number of samples,  \( \text{acc}(B_m) \) is the empirical accuracy of bin \( m \),  and \( \text{conf}(B_m) \) is the average confidence in bin \( m \).   

Although widely adopted, ECE is sensitive to the choice of binning and may smooth over severe local miscalibration. For uncertainty calibration, we define confidence as (\( 1 - \) uncertainty score) for uncertainty measures where higher values indicate lower confidence (e.g., entropy-based uncertainty). For other metrics where higher values indicate greater confidence, we directly use the uncertainty score.



\textbf{Maximum Expected Calibration Error (MECE):}~\cite{guo2017calibration} captures the worst-case miscalibration, addressing binning artifacts in ECE:
\begin{equation}
\mathit{MECE} = \max_{m \in M} \left| \text{acc}(B_m) - \text{conf}(B_m) \right|
\end{equation}
Unlike ECE, MECE highlights peak miscalibration, making it more suitable when even small pockets of overconfidence can be risky. However, it is highly sensitive to outliers.
\textbf{Brier Score (BS)} evaluates the mean squared error between predicted probabilities and ground truth labels: 
\begin{equation}
B = \frac{1}{|T|} \sum_{i=1}^{|T|} (\hat{p}_i - (\checkmark(y_i)))^2
\end{equation}

where \( |T| \) is the total number of samples,  \( \hat{p}_i \) is the predicted probability for sample \( i \),  \( (\checkmark(y_i)) \) is an indicator function that is 1 if the prediction is correct, 0 otherwise. It penalizes both miscalibration and incorrect predictions, but it cannot disentangle whether errors stem from poor uncertainty estimates or low accuracy. 


To support selective prediction we use the \textbf{area under the accuracy-rejection curve (AUARC)}~\cite{pmlr-v8-nadeem10a}, which is specifically designed for uncertainty estimation, which plots how the accuracy on the accepted samples changes as more low-confidence answers are rejected. The area under it indicates the uncertainty estimation’s ability to maintain high accuracy when low-confidence answers are rejected. AUARC measures how well uncertainty scores support abstention but can exaggerate performance when abstention primarily removes trivial, highly uncertain failures. 

We conducted all our experiments on a server equipped with an Intel(R) Xeon(R) w5-2545 CPU and two RTX 4000 Ada GPUs. Due to space limitation complete tables associated on AUARC values are provided in our project website.

\begin{figure*}[!t]
    \centering

    \begin{subfigure}[t]{0.24\textwidth}
        \centering
        \includegraphics[width=\linewidth]{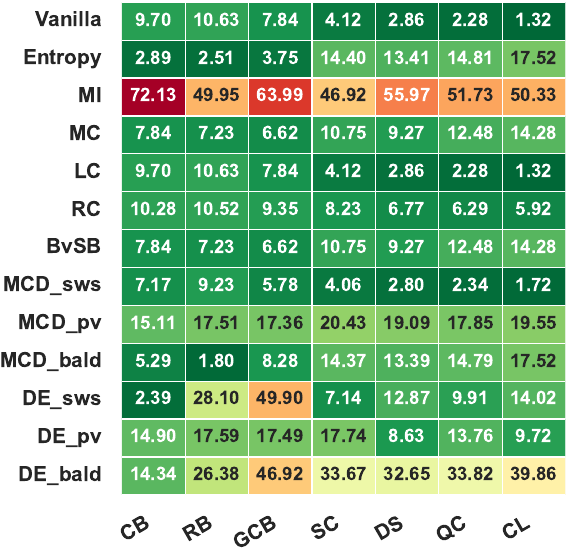}
        \caption{Uncalibrated}
        \label{dp1}
    \end{subfigure}\hfill
    \begin{subfigure}[t]{0.24\textwidth}
        \centering
        \includegraphics[width=\linewidth]{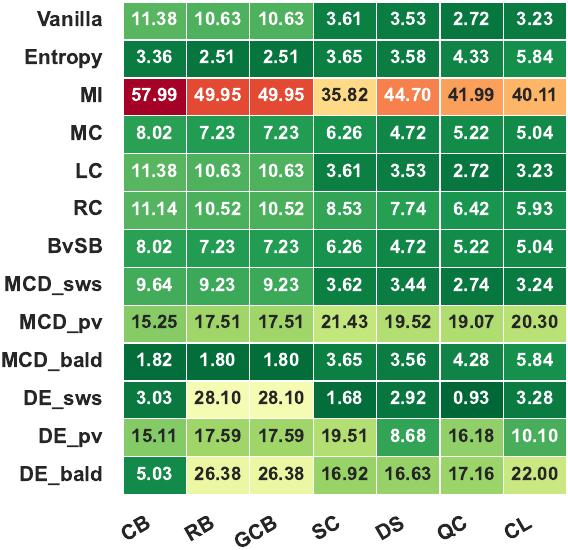}
        \caption{Temperature Scaling}
        \label{dp2}
    \end{subfigure}\hfill
    \begin{subfigure}[t]{0.24\textwidth}
        \centering
        \includegraphics[width=\linewidth]{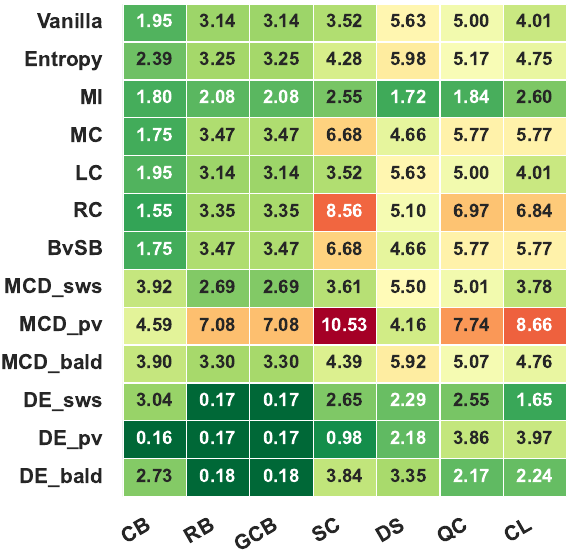}
        \caption{Platt Scaling}
        \label{dp3}
    \end{subfigure}\hfill
    \begin{subfigure}[t]{0.24\textwidth}
        \centering
        \includegraphics[width=\linewidth]{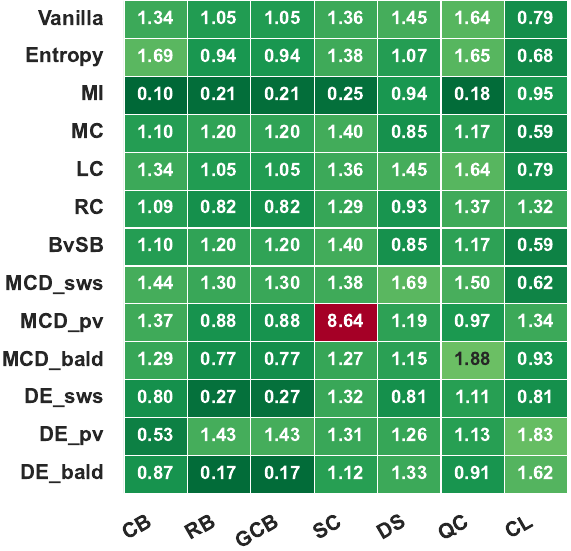}
        \caption{Isotonic Regression}
        \label{dp4}
    \end{subfigure}\hfill

    \caption{ECE values on defect prediction task for CodeBERT (CB), RoBERT (RB), GraphCodeBERT (GCB), StarCoder (SC), DeepSeek (DS), QwenCoder (QC), and CodeLlama (CL) models}
\end{figure*}

\begin{figure*}[!t]
    \centering

    \begin{subfigure}[t]{0.24\textwidth}
        \centering
        \includegraphics[width=\linewidth]{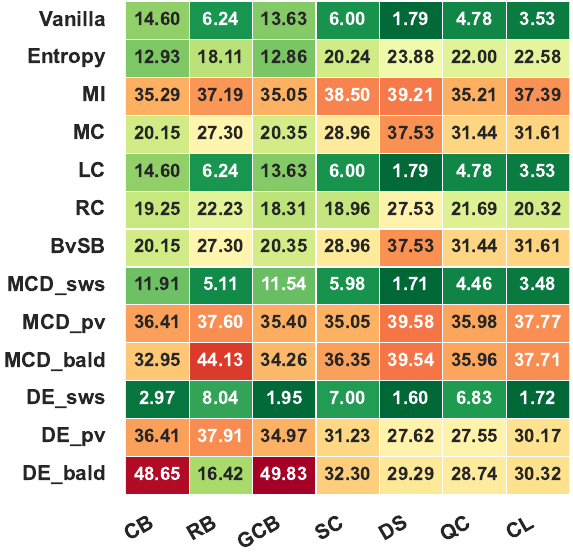}
        \caption{Uncalibrated}
        \label{vp1}
    \end{subfigure}\hfill
    \begin{subfigure}[t]{0.24\textwidth}
        \centering
        \includegraphics[width=\linewidth]{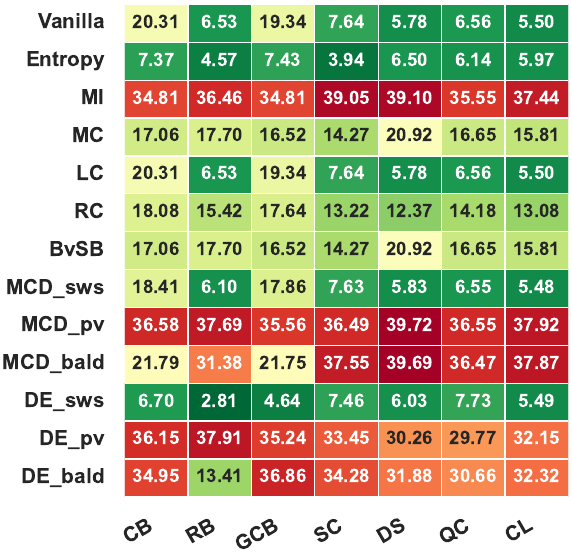}
        \caption{Temperature Scaling}
        \label{vp2}
    \end{subfigure}\hfill
    \begin{subfigure}[t]{0.24\textwidth}
        \centering
        \includegraphics[width=\linewidth]{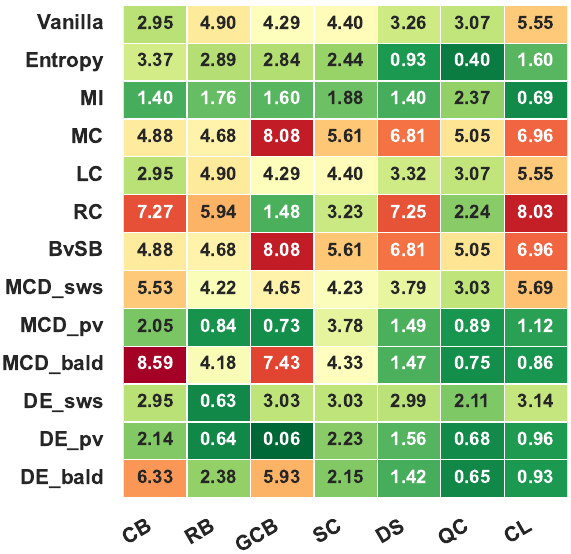}
        \caption{Platt Scaling}
        \label{vp3}
    \end{subfigure}\hfill
    \begin{subfigure}[t]{0.24\textwidth}
        \centering
        \includegraphics[width=\linewidth]{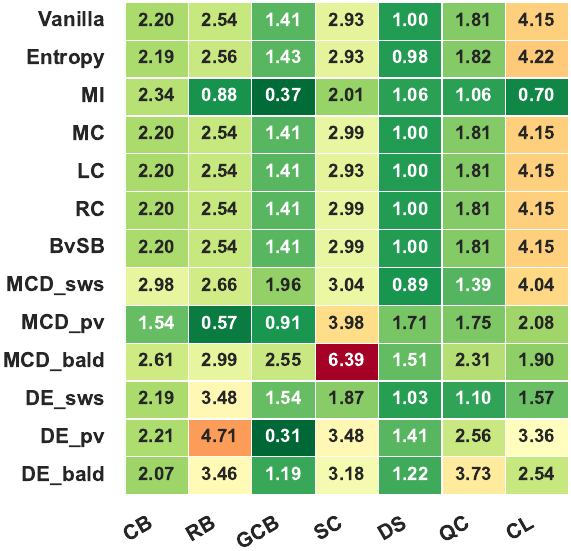}
        \caption{Isotonic Regression}
        \label{vp4}
    \end{subfigure}\hfill

    \caption{ECE values on vulnerability detection task for CodeBERT (CB), RoBERT (RB), GraphCodeBERT (GCB), StarCoder (SC), DeepSeek (DS), QwenCoder (QC), and CodeLlama (CL) models}
\end{figure*}

\section{Empirical Study }
\label{sec::emperical}

This empirical study addresses two objectives. First, it evaluates whether commonly used prediction scores and uncertainty estimates provide effective signals for \textbf{selective prediction}~\cite{xin2021art} in code classification. Second, it examines whether standard post-hoc calibration methods improve the usefulness of these signals for selective prediction.

To this end, we conduct experiments on deep code models for two software engineering tasks: vulnerability detection and defect prediction. Using the experimental setup described in Section~\ref{sec::methodology}, we investigate the following research questions:

\textbf{RQ1.}  How well do existing uncertainty metrics reflect prediction reliability and calibration quality in deep code models?



\textbf{RQ2.} Can existing calibration techniques improve the effectiveness of these measurement metrics in determining input appropriateness?

\textbf{RQ3.} Do commonly used calibration metrics reliably indicate an uncertainty measure’s ability to assess prediction success in deep learning models?




\subsection{Uncalibrated Reliability of Uncertainty Measures}

Tables~\ref{vd_mece}--\ref{bs_dp} and Figures~\ref{dp1}--\ref{vp1} summarize the \emph{uncalibrated} reliability of uncertainty signals across our subjects. In software engineering settings, an uncertainty signal can remain \emph{useful for ranking} while being \emph{unsafe as a probability}. This gap is large, task-dependent, and directly relevant to deployment decisions.

\textbf{Uncalibrated uncertainty is not a dependable measure in SE tasks.}
Before calibration, the numeric mismatch between confidence and correctness is substantial. In vulnerability detection, ECE ranges from \textbf{1.79\% to 48.65\%} while in defect prediction it is even larger, reaching \textbf{72.13\%}. The same pattern appears when MECE rises to \textbf{65.8}, while BS is consistently lower for probability-based confidence scores (around \(0.225\)--
\(0.244\) on vulnerability detection and around \( 0.125\)--\(0.157\) on defect prediction) than for information-theoretic signals (\(0.360\)--\(0.397\) and \(0.460\)--\(0.699\), respectively). This is not merely generic neural overconfidence. In code tasks, correctness often depends on semantics, hidden interactions, and long-range relations, whereas models can still assign high confidence from lexical or syntactic regularities. Consequently, score magnitude often reflects \emph{pattern familiarity} rather than true correctness.

\textbf{Ranking reliability and probability reliability diverge in code modeling.}
Despite severe miscalibration, AUARC remains relatively strong for many signals: \textbf{0.681--0.754} in vulnerability detection and \textbf{0.885--0.927} in defect prediction. For example, in vulnerability detection, CodeBERT attains AUARC near \textbf{0.754} across several signals even when ECE varies from \textbf{11.91\%} to above \textbf{35\%}. This shows that, for SE tasks, a signal may still separate easy from risky inputs well enough for abstention while failing to quantify how trustworthy the retained predictions actually are. Hence, \textbf{good rejection behavior does not imply trustworthy probabilities}. This distinction matters in practice when ranking can help triage predictions for review, but uncalibrated confidence is unsafe for threshold-based automation.

\textbf{No uncertainty measurement metric is generalizable across SE tasks and code models.}
No single uncertainty measure is consistently reliable across both tasks and all architectures. The same metric can be well behaved for one model yet highly unreliable for another, and this pattern changes between vulnerability detection and defect prediction. Information-theoretic measures illustrate this most clearly as they often incur substantially higher ECE and BS, and weaker pre-calibration AUARC, than confidence-based alternatives. This suggests that uncertainty behavior is shaped by \textbf{task-specific code distributions}, \textbf{label ambiguity}, and \textbf{model-specific inductive biases}, rather than by the metric alone. Therefore, there is no universally reliable off-the-shelf uncertainty score for SE deployment.

\textbf{Takeaway. }
Existing uncertainty measures in deep code models are best viewed as \textbf{task-specific ranking heuristics, not trustworthy probabilities}. Their failure arises from SE-specific error structure---semantic ambiguity, overlap in surface patterns across classes, noisy supervision, and missed long-range program dependencies---which weakens the link between confidence magnitude and correctness. For practitioners, the implication is straightforward: \textbf{do not interpret uncalibrated confidence as reliability}. Instead, validate each signal per task and model, and report \emph{both} calibration metrics and AUARC before using uncertainty to support acceptance, deferral, or review prioritization.

\begin{figure}[t]
    \centering

    \begin{subfigure}[t]{0.49\columnwidth}
        \centering
        \includegraphics[width=\linewidth]{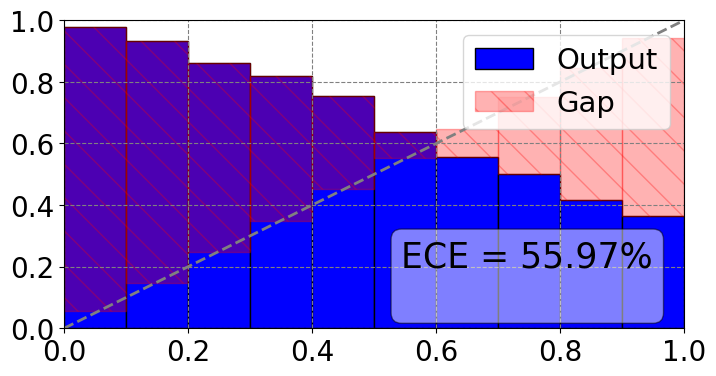}
        \caption{Reliability plot of uncalibrated model}
        \label{MI1}
    \end{subfigure}\hfill
    \begin{subfigure}[t]{0.49\columnwidth}
        \centering
        \includegraphics[width=\linewidth]{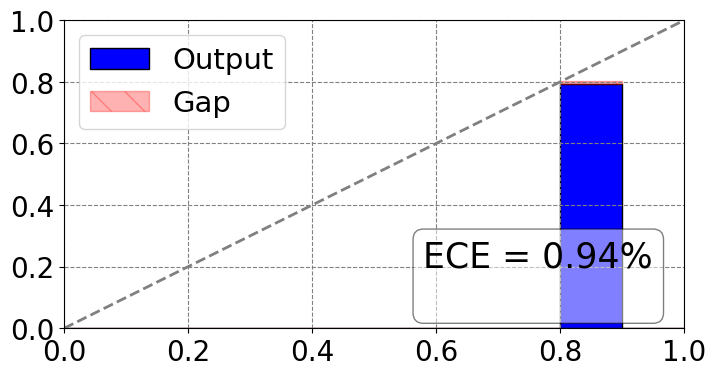}
        \caption{Reliability plot showing bucket collapse with Iso.R}
        \label{MI2}
    \end{subfigure}\hfill

    \caption{Example of bucket collapse for MI after Iso.R on defect prediction with DeepSeek.}
    \label{bucket}
\end{figure}

\subsection{Effectiveness of Calibration Techniques on Uncertainty Metrics }

Tables~\ref{vd_mece}--\ref{bs_dp} and Figures~\ref{dp2}--\ref{dp4} and \ref{vp2}--\ref{vp4} summarize the effect of post-hoc calibration on uncertainty signals. For SE tasks, calibration often improves \emph{probability alignment} without improving \emph{decision utility}. That is, a calibrated score may look numerically better while remaining no more useful for deciding which predictions to trust, defer, or inspect.

\textbf{Monotone calibration preserves ordering but changes probability interpretation.}
This distinction is important when interpreting the effect of calibration on selective prediction. Let \(s_i\) and \(s_j\) denote the scores used to rank predictions for rejection, and let \(f\) be a strictly monotone calibration map. Then \(s_i > s_j\) if and only if \(f(s_i) > f(s_j)\). Therefore, at any fixed coverage level, the set of accepted predictions is unchanged by applying \(f\). Ranking-based selective-prediction metrics such as AUARC are consequently invariant to strictly monotone calibration, except for implementation-level effects such as ties, non-strict mappings, or numerical precision. Calibration metrics and selective-prediction metrics should therefore be interpreted as measuring different properties: ECE and BS assess probability validity, whereas AUARC assesses the quality of the accept--defer ranking.

\textbf{Post-hoc calibration improves probability alignment without necessarily improving failure separation.}
Across models and tasks, PS and Iso.R usually reduce ECE sharply. In vulnerability detection, PS often brings ECE down to \textbf{0.403\%--8.587\%}, while Iso.R frequently reaches \textbf{0.880\%--4.223\%}. In defect prediction, both methods commonly reduce ECE to \textbf{below 10\%}, and in several cases to \textbf{0.10\%--0.95\%}. In contrast, TS is unstable and can even worsen calibration; for example, Vanilla confidence for CodeBERT increases from \textbf{14.596\% to 20.314\%} in vulnerability detection and from \textbf{9.70\% to 11.38\%} in defect prediction. This behavior is important in SE because code-model errors are not homogeneous, thus they arise from semantically subtle cases, subgroup-specific patterns, and long-range dependencies. A single global remapping can therefore improve average alignment while still missing the error structure that matters in code analysis.

\begin{table*}[t]
\centering
\caption{MECE values on vulnerability detection task. Each entry shows uncalibrated $\rightarrow$ best calibrated technique in terms of MECE. Superscripts denote which calibration achieved the best value.}
\label{vd_mece}
\renewcommand{\arraystretch}{}
\resizebox{0.8\linewidth}{!}{
\begin{tabular}{llllllll}
\toprule
\textbf{Metric} & \textbf{CodeBERT} & \textbf{RoBERTa} & \textbf{GraphCodeBERT} & \textbf{StarCoder} & \textbf{DeepSeek} & \textbf{QwenCoder} & \textbf{CodeLlama} \\
\midrule
Vanilla   & 24.10$\rightarrow$12.50$^{Iso.R}$ & 19.00$\rightarrow$9.80$^{PS}$ & 25.30$\rightarrow$7.10$^{PS}$ & 13.10$\rightarrow$13.90$^{TS}$ & 4.00$\rightarrow$8.00$^{TS}$ & 11.70$\rightarrow$11.80$^{TS}$ & 8.20$\rightarrow$8.60$^{PS}$ \\
Entropy   & 19.50$\rightarrow$5.40$^{Iso.R}$  & 23.30$\rightarrow$8.30$^{TS}$ & 20.20$\rightarrow$6.90$^{PS}$ & 21.50$\rightarrow$7.50$^{PS}$  & 33.00$\rightarrow$2.70$^{PS}$ & 25.70$\rightarrow$1.30$^{PS}$  & 28.10$\rightarrow$3.90$^{PS}$ \\
MI        & \textbf{52.20$\rightarrow$2.30$^{Iso.R}$}  & 59.50$\rightarrow$0.90$^{Iso.R}$ & 51.90$\rightarrow$0.40$^{Iso.R}$ & \textbf{65.80$\rightarrow$2.00$^{Iso.R}$}  & \textbf{55.40$\rightarrow$1.10$^{Iso.R}$} & \textbf{64.80$\rightarrow$1.10$^{Iso.R}$} & \textbf{60.10$\rightarrow$0.70$^{Iso.R}$} \\
MC        & 45.70$\rightarrow$12.50$^{Iso.R}$ & 49.30$\rightarrow$47.90$^{PS}$ & 48.50$\rightarrow$83.30$^{PS}$ & 46.30$\rightarrow$47.90$^{TS}$  & 46.90$\rightarrow$27.60$^{Iso.R}$ & 51.00$\rightarrow$16.70$^{Iso.R}$ & 46.30$\rightarrow$39.70$^{Iso.R}$ \\
RC        & 44.40$\rightarrow$12.50$^{Iso.R}$ & 46.60$\rightarrow$49.00$^{PS}$ & 50.30$\rightarrow$64.10$^{PS}$ & 45.20$\rightarrow$47.80$^{PS}$  & 46.90$\rightarrow$27.60$^{Iso.R}$ & 52.30$\rightarrow$16.70$^{Iso.R}$ & 45.90$\rightarrow$39.70$^{Iso.R}$ \\
MCD\_sws  & 15.50$\rightarrow$6.80$^{PS}$   & 13.80$\rightarrow$10.10$^{PS}$ & 17.10$\rightarrow$8.70$^{PS}$ & 13.30$\rightarrow$12.40$^{Iso.R}$ & 4.00$\rightarrow$7.50$^{TS}$  & 11.50$\rightarrow$11.30$^{TS}$ & 8.20$\rightarrow$8.10$^{TS}$ \\
MCD\_pv   & 36.40$\rightarrow$29.80$^{PS}$  & 37.60$\rightarrow$18.00$^{Iso.R}$ & 35.40$\rightarrow$6.00$^{Iso.R}$  & 62.20$\rightarrow$10.30$^{Iso.R}$ & 80.80$\rightarrow$11.40$^{PS}$ & 86.80$\rightarrow$77.10$^{TS}$ & 37.80$\rightarrow$37.90$^{TS}$ \\
MCD\_bald & 46.70$\rightarrow$35.00$^{Iso.R}$ & 53.50$\rightarrow$16.00$^{Iso.R}$ & 50.40$\rightarrow$20.20$^{Iso.R}$ & 51.90$\rightarrow$16.30$^{Iso.R}$ & 81.00$\rightarrow$24.50$^{PS}$ & 72.50$\rightarrow$60.90$^{TS}$ & 37.80$\rightarrow$9.00$^{PS}$ \\
DE\_sws   & 3.70$\rightarrow$9.30$^{PS}$    & 8.00$\rightarrow$0.60$^{PS}$ & 5.20$\rightarrow$7.30$^{TS}$  & 19.70$\rightarrow$11.20$^{Iso.R}$ & 17.30$\rightarrow$8.00$^{Iso.R}$ & 15.80$\rightarrow$16.90$^{PS}$ & 11.10$\rightarrow$6.00$^{TS}$ \\
DE\_pv    & 36.40$\rightarrow$19.30$^{PS}$   & \textbf{37.90$\rightarrow$0.60$^{PS}$} & \textbf{35.00$\rightarrow$0.10$^{PS}$} & 100.00$\rightarrow$14.70$^{Iso.R}$ & 100.00$\rightarrow$49.50$^{Iso.R}$ & 54.40$\rightarrow$6.80$^{Iso.R}$ & 39.50$\rightarrow$4.70$^{PS}$ \\
DE\_bald  & 52.60$\rightarrow$21.20$^{Iso.R}$ & 52.80$\rightarrow$14.90$^{Iso.R}$ & 59.90$\rightarrow$38.50$^{PS}$ & 36.90$\rightarrow$37.50$^{TS}$  & 100.00$\rightarrow$19.00$^{PS}$ & 42.80$\rightarrow$7.80$^{Iso.R}$ & 32.20$\rightarrow$11.30$^{PS}$ \\
\bottomrule
\end{tabular}

}
\end{table*}

\begin{table*}[t]
\centering
\renewcommand{\arraystretch}{}
\caption{MECE values on defect prediction task. Each entry shows uncalibrated $\rightarrow$ best calibrated technique in terms of MECE. Superscripts denote which calibration achieved the best value.}
\label{dp_mece}
\resizebox{0.8\linewidth}{!}{
\begin{tabular}{llllllll}
\toprule
\textbf{Metric} & \textbf{CodeBERT} & \textbf{RoBERTa} & \textbf{GraphCodeBERT} & \textbf{StarCoder} & \textbf{DeepSeek} & \textbf{QwenCoder} & \textbf{CodeLlama} \\
\midrule
Vanilla & 37.69$\rightarrow$13.10$^{Iso.R}$ & 20.72$\rightarrow$21.94$^{TS}$ & 20.72$\rightarrow$21.94$^{TS}$ & 29.60$\rightarrow$5.65$^{TS}$ & 29.14$\rightarrow$24.29$^{Iso.R}$ & 29.14$\rightarrow$17.55$^{PS}$ & 9.06$\rightarrow$14.06$^{Iso.R}$ \\
Entropy & 49.02$\rightarrow$11.76$^{Iso.R}$ & 49.34$\rightarrow$41.08$^{PS}$ & 49.34$\rightarrow$41.08$^{PS}$ & 27.95$\rightarrow$18.88$^{TS}$ & 30.82$\rightarrow$20.90$^{TS}$ & 29.39$\rightarrow$9.47$^{Iso.R}$ & 28.22$\rightarrow$13.77$^{Iso.R}$ \\
MI & \textbf{88.23$\rightarrow$0.10$^{Iso.R}$} & 88.35$\rightarrow$0.21$^{Iso.R}$ & 88.35$\rightarrow$0.21$^{Iso.R}$ & \textbf{89.51$\rightarrow$0.25$^{Iso.R}$} & \textbf{92.24$\rightarrow$0.94$^{Iso.R}$} & \textbf{92.14$\rightarrow$0.18$^{Iso.R}$} & 91.80$\rightarrow$16.13$^{PS}$ \\
MC & 42.96$\rightarrow$7.64$^{Iso.R}$ & 35.63$\rightarrow$10.56$^{Iso.R}$ & 35.63$\rightarrow$10.56$^{Iso.R}$ & 37.93$\rightarrow$6.17$^{Iso.R}$ & 38.73$\rightarrow$3.70$^{Iso.R}$ & 38.37$\rightarrow$12.23$^{Iso.R}$ & 35.21$\rightarrow$2.40$^{Iso.R}$ \\
RC & 35.68$\rightarrow$27.78$^{Iso.R}$ & 34.42$\rightarrow$8.47$^{Iso.R}$ & 34.42$\rightarrow$8.47$^{Iso.R}$ & 36.68$\rightarrow$3.15$^{Iso.R}$ & 38.11$\rightarrow$6.69$^{Iso.R}$ & 37.82$\rightarrow$13.19$^{Iso.R}$ & 36.33$\rightarrow$3.22$^{Iso.R}$ \\
MCD\_sws & 48.87$\rightarrow$23.55$^{PS}$ & 14.98$\rightarrow$25.43$^{TS}$ & 14.98$\rightarrow$25.43$^{TS}$ & 29.27$\rightarrow$5.74$^{TS}$ & 28.90$\rightarrow$24.60$^{Iso.R}$ & 28.96$\rightarrow$7.92$^{Iso.R}$ & 11.60$\rightarrow$16.55$^{PS}$ \\
MCD\_pv & 15.11$\rightarrow$4.57$^{Iso.R}$ & 17.36$\rightarrow$2.39$^{Iso.R}$ & 17.36$\rightarrow$2.39$^{Iso.R}$ & 82.70$\rightarrow$58.74$^{PS}$ & 100.00$\rightarrow$4.83$^{Iso.R}$ & 71.51$\rightarrow$16.67$^{Iso.R}$ & 49.03$\rightarrow$44.47$^{PS}$ \\
MCD\_bald & 26.07$\rightarrow$27.99$^{TS}$ & 43.72$\rightarrow$41.06$^{PS}$ & 43.72$\rightarrow$41.06$^{PS}$ & 27.75$\rightarrow$18.71$^{TS}$ & 30.01$\rightarrow$24.45$^{TS}$ & 27.51$\rightarrow$15.93$^{Iso.R}$ & 27.82$\rightarrow$11.42$^{Iso.R}$ \\
DE\_sws & 26.54$\rightarrow$36.65$^{TS}$ & 49.91$\rightarrow$2.20$^{PS}$ & 49.91$\rightarrow$2.20$^{PS}$ & 13.18$\rightarrow$1.77$^{Iso.R}$ & 16.96$\rightarrow$3.62$^{PS}$ & 16.69$\rightarrow$16.01$^{Iso.R}$ & \textbf{18.26$\rightarrow$1.93$^{Iso.R}$} \\
DE\_pv & 14.90$\rightarrow$3.87$^{PS}$ & 17.49$\rightarrow$0.17$^{PS}$ & 
17.49$\rightarrow$0.17$^{PS}$ & 19.04$\rightarrow$17.79$^{Iso.R}$ & 94.10$\rightarrow$30.00$^{Iso.R}$ & 100.00$\rightarrow$81.10$^{PS}$ & 100.00$\rightarrow$50.00$^{Iso.R}$ \\
DE\_bald & 53.11$\rightarrow$17.84$^{Iso.R}$ & \textbf{74.70$\rightarrow$0.18$^{PS}$} & \textbf{74.70$\rightarrow$0.18$^{PS}$} & 49.48$\rightarrow$34.39$^{TS}$ & 47.87$\rightarrow$13.14$^{Iso.R}$ & 50.45$\rightarrow$16.67$^{Iso.R}$ & 52.64$\rightarrow$40.47$^{TS}$ \\
\bottomrule
\end{tabular}
}
\end{table*}

\textbf{Better calibration scores rarely imply better selective prediction.}
Despite large reductions in ECE and BS, selective prediction performance usually improves only marginally. In vulnerability detection, CodeBERT remains near \textbf{0.754} across several calibration settings even when ECE falls to roughly \textbf{2--3\%}. In defect prediction, AUARC for many metrics remains around \textbf{0.886--0.928} even when ECE drops close to \textbf{1\%}. Thus, calibration can improve the \emph{numerical meaning} of a score without improving its \emph{ability to separate correct from incorrect predictions}. For SE practitioners, this is the key warning when a lower ECE does not mean the model has become safer to automate. It may still retain high-confidence wrong predictions, which are precisely the failures that matter most in vulnerability screening and defect triage.

\textbf{The main limitation of post-hoc calibration is that it improves average probability alignment without correcting the error distribution that governs which code predictions remain risky after thresholding.}
TS, PS, and Iso.R are all optimized to improve global agreement between predicted scores and observed correctness, rather than to separate correct from incorrect predictions among the instances the model is most likely to retain. This becomes explicit in defect prediction, where Iso.R reduces ECE for MI-based scores to \textbf{0.10\%--0.95\%}, yet AUARC remains only \textbf{0.772--0.843}, well below the \textbf{0.915--0.928} reached by stronger confidence-based signals. The root cause is particularly relevant in code tasks where model mistakes are often concentrated in semantically hard or distributionally sparse cases, whereas post-hoc calibration learns a single monotone mapping over the full score range. As a result, calibration can fix \emph{score scale} while leaving \textbf{high-confidence errors interleaved with correct predictions}. Therefore, calibration and uncertainty must be treated jointly in SE, not as independent post-processing steps.

\textbf{Takeaway.}
Standard post-hoc calibration techniques improve numeric calibration in deep code models, but they do \textbf{not} reliably improve practical reliability for SE deployment. Their failure is not incidental as it follows from a mismatch between what calibration optimizes is the global alignment and what SE applications require is safe trust decisions in the high-confidence region. For practitioners, the implication is to \textbf{use calibration to refine probability interpretation, not to assume better abstention behavior}. A calibrated uncertainty score should be trusted for acceptance or deferral after verifying that selective prediction metrics also improve.

\begin{table*}[t]
\centering
\caption{BS values on defect prediction. Each entry shows uncalibrated $\rightarrow$ best calibrated technique in terms of BS. Superscripts denote which calibration achieved the best value.}
\label{bs_vd}
\renewcommand{\arraystretch}{}
\resizebox{0.8\linewidth}{!}{
\begin{tabular}{llllllll}
\toprule
\textbf{Metric} & \textbf{CodeBERT} & \textbf{RoBERTa} & \textbf{GraphCodeBERT} & \textbf{StarCoder} & \textbf{DeepSeek} & \textbf{QwenCoder} & \textbf{CodeLlama} \\
\midrule
Vanilla   & 24.41$\rightarrow$21.73$^{Iso.R}$ & 22.98$\rightarrow$22.39$^{Iso.R}$ & 23.96$\rightarrow$21.65$^{Iso.R}$ & 23.80$\rightarrow$23.42$^{Iso.R}$ & 22.66$\rightarrow$22.74$^{Iso.R}$ & 23.11$\rightarrow$22.83$^{Iso.R}$ & 22.51$\rightarrow$22.50$^{Iso.R}$ \\
Entropy   & 23.67$\rightarrow$21.72$^{Iso.R}$ & 25.98$\rightarrow$22.37$^{PS}$  & 23.67$\rightarrow$21.66$^{Iso.R}$ & 27.52$\rightarrow$23.37$^{PS}$  & \underline{28.51$\rightarrow$22.62$^{PS}$} & 27.81$\rightarrow$22.75$^{PS}$ & 27.55$\rightarrow$22.35$^{PS}$ \\
MI        & 36.72$\rightarrow$21.76$^{PS}$  & \underline{37.42$\rightarrow$22.22$^{PS}$} & 36.06$\rightarrow$21.54$^{PS}$ & \underline{39.44$\rightarrow$23.29$^{PS}$} & 39.66$\rightarrow$22.64$^{PS}$ & \underline{36.99$\rightarrow$22.74$^{PS}$} & \underline{37.64$\rightarrow$22.29$^{PS}$} \\
MC        & 27.50$\rightarrow$21.73$^{Iso.R}$ & 32.74$\rightarrow$22.39$^{Iso.R}$ & 27.37$\rightarrow$21.65$^{Iso.R}$ & 33.73$\rightarrow$23.43$^{Iso.R}$ & 37.62$\rightarrow$22.74$^{Iso.R}$ & 34.86$\rightarrow$22.83$^{Iso.R}$ & 34.03$\rightarrow$22.50$^{Iso.R}$ \\
RC        & 26.44$\rightarrow$21.73$^{Iso.R}$ & 29.21$\rightarrow$22.39$^{Iso.R}$ & 26.19$\rightarrow$21.65$^{Iso.R}$ & 29.19$\rightarrow$23.43$^{Iso.R}$ & 31.94$\rightarrow$22.74$^{Iso.R}$ & 30.18$\rightarrow$22.83$^{Iso.R}$ & 29.00$\rightarrow$22.50$^{Iso.R}$ \\
MCD\_sws  & 22.99$\rightarrow$21.58$^{Iso.R}$ & 22.68$\rightarrow$22.43$^{Iso.R}$ & 22.98$\rightarrow$21.56$^{Iso.R}$ & 23.79$\rightarrow$23.41$^{Iso.R}$ & 22.66$\rightarrow$22.75$^{Iso.R}$ & 23.10$\rightarrow$22.87$^{Iso.R}$ & 22.52$\rightarrow$22.56$^{Iso.R}$ \\
MCD\_pv   & 36.49$\rightarrow$21.86$^{Iso.R}$ & 37.66$\rightarrow$22.74$^{Iso.R}$ & 35.47$\rightarrow$21.53$^{Iso.R}$ & 36.22$\rightarrow$23.60$^{Iso.R}$ & 39.66$\rightarrow$23.78$^{Iso.R}$ & 36.42$\rightarrow$23.09$^{Iso.R}$ & 37.87$\rightarrow$23.17$^{Iso.R}$ \\
MCD\_bald & \underline{35.33$\rightarrow$21.57$^{Iso.R}$} & 43.75$\rightarrow$22.36$^{Iso.R}$ & 35.91$\rightarrow$21.54$^{Iso.R}$ & 37.22$\rightarrow$24.13$^{Iso.R}$ & 39.65$\rightarrow$24.00$^{Iso.R}$ & 36.37$\rightarrow$23.25$^{Iso.R}$ & 37.84$\rightarrow$23.57$^{Iso.R}$ \\
DE\_sws   & 21.71$\rightarrow$21.72$^{Iso.R}$ & 24.18$\rightarrow$23.47$^{Iso.R}$ & 21.40$\rightarrow$21.49$^{PS}$  & 24.17$\rightarrow$23.63$^{Iso.R}$ & 23.40$\rightarrow$23.43$^{Iso.R}$ & 23.69$\rightarrow$23.16$^{Iso.R}$ & 23.14$\rightarrow$23.16$^{Iso.R}$ \\
DE\_pv    & 36.49$\rightarrow$23.32$^{Iso.R}$ & 37.91$\rightarrow$23.55$^{PS}$  & 35.32$\rightarrow$22.79$^{Iso.R}$ & 34.62$\rightarrow$24.01$^{Iso.R}$ & 32.36$\rightarrow$24.07$^{PS}$  & 31.61$\rightarrow$23.17$^{Iso.R}$ & 32.79$\rightarrow$23.53$^{PS}$ \\
DE\_bald  & 46.08$\rightarrow$21.75$^{Iso.R}$ & 27.33$\rightarrow$23.47$^{Iso.R}$ & \underline{47.38$\rightarrow$21.44$^{Iso.R}$} & 35.24$\rightarrow$24.20$^{PS}$ & 33.22$\rightarrow$24.01$^{PS}$ & 32.19$\rightarrow$23.50$^{Iso.R}$ & 32.99$\rightarrow$23.54$^{PS}$ \\
\bottomrule
\end{tabular}
}

\end{table*}

\begin{table*}[t]
\centering
\caption{BS values on vulnerability detection. Each entry shows uncalibrated $\rightarrow$ best calibrated technique in terms of BS. Superscripts denote which calibration achieved the best value.}
\label{bs_dp}
\renewcommand{\arraystretch}{}
\resizebox{0.8\linewidth}{!}{
\begin{tabular}{llllllll}
\toprule
\textbf{Metric} & \textbf{CodeBERT} & \textbf{RoBERTa} & \textbf{GraphCodeBERT} & \textbf{StarCoder} & \textbf{DeepSeek} & \textbf{QwenCoder} & \textbf{CodeLlama} \\
\midrule
Vanilla   & $12.50\rightarrow 11.30^{Iso.R}$ & $13.30\rightarrow 12.70^{Iso.R}$ & $13.30\rightarrow 12.70^{Iso.R}$ & $\underline{15.70\rightarrow 15.50^{Iso.R}}$ & $13.00\rightarrow 13.00^{TS/Iso.R}$ & $\underline{13.70\rightarrow 13.70^{TS/Iso.R}}$ & $\underline{13.50\rightarrow 13.50^{Iso.R}}$ \\
Entropy   & $11.30\rightarrow 11.30^{TS/Iso.R}$ & $12.90\rightarrow 12.70^{TS/Iso.R}$ & $12.90\rightarrow 12.70^{TS/Iso.R}$ & $18.30\rightarrow 15.60^{Iso.R}$ & $15.30\rightarrow 13.00^{Iso.R}$ & $16.70\rightarrow 13.90^{Iso.R}$ & $17.40\rightarrow 13.60^{Iso.R}$ \\
MI        & $69.90\rightarrow 13.20^{Iso.R}$ & $62.10\rightarrow 14.70^{Iso.R}$ & $62.10\rightarrow 14.70^{Iso.R}$ & $46.00\rightarrow 18.30^{Iso.R}$ & $53.70\rightarrow 16.40^{Iso.R}$ & $50.50\rightarrow 17.20^{Iso.R}$ & $48.70\rightarrow 17.30^{Iso.R}$ \\
MC        & $12.20\rightarrow 11.30^{Iso.R}$ & $13.70\rightarrow 12.70^{PS/Iso.R}$ & $13.70\rightarrow 12.70^{PS/Iso.R}$ & $18.30\rightarrow 15.70^{TS/Iso.R}$ & $15.20\rightarrow 13.00^{Iso.R}$ & $16.90\rightarrow 13.80^{Iso.R}$ & $17.20\rightarrow 13.80^{TS/Iso.R}$ \\
RC        & $12.70\rightarrow 11.40^{Iso.R}$ & $13.80\rightarrow 12.70^{TS/Iso.R}$ & $13.80\rightarrow 12.70^{TS/Iso.R}$ & $16.90\rightarrow 15.70^{Iso.R}$ & $14.10\rightarrow 13.10^{Iso.R}$ & $14.90\rightarrow 13.90^{Iso.R}$ & $14.80\rightarrow 13.90^{Iso.R}$ \\
MCD\_sws  & $11.80\rightarrow 11.10^{Iso.R}$ & $12.90\rightarrow 12.60^{PS/Iso.R}$ & $12.90\rightarrow 12.60^{PS/Iso.R}$ & $\underline{15.70\rightarrow 15.50^{Iso.R}}$ & $13.00\rightarrow 13.00^{TS/Iso.R}$ & $\underline{13.70\rightarrow 13.70^{TS/Iso.R}}$ & $\underline{13.50\rightarrow 13.50^{Iso.R}}$ \\
MCD\_pv   & $15.20\rightarrow 11.20^{Iso.R}$ & $17.50\rightarrow 12.60^{Iso.R}$ & $17.50\rightarrow 12.60^{Iso.R}$ & $21.60\rightarrow 17.50^{Iso.R}$ & $19.60\rightarrow 14.20^{Iso.R}$ & $19.40\rightarrow 15.10^{Iso.R}$ & $20.40\rightarrow 15.30^{Iso.R}$ \\
MCD\_bald & $\underline{11.50\rightarrow 11.00^{TS}}$ & $\underline{13.60\rightarrow 12.50^{TS/Iso.R}}$ & $\underline{13.60\rightarrow 12.50^{TS/Iso.R}}$ & $18.30\rightarrow 15.60^{Iso.R}$ & $\underline{15.30\rightarrow 12.90^{Iso.R}}$ & $16.70\rightarrow 13.90^{Iso.R}$ & $17.40\rightarrow 13.60^{Iso.R}$ \\
DE\_sws   & $12.40\rightarrow 12.30^{Iso.R}$ & $39.60\rightarrow 14.70^{PS/Iso.R}$ & $39.60\rightarrow 14.70^{PS/Iso.R}$ & $17.70\rightarrow 17.00^{TS/Iso.R}$ & $16.70\rightarrow 14.90^{Iso.R}$ & $17.30\rightarrow 16.10^{TS/Iso.R}$ & $18.00\rightarrow 15.80^{PS/Iso.R}$ \\
DE\_pv    & $15.30\rightarrow 12.60^{Iso.R}$ & $17.70\rightarrow 14.60^{PS}$ & $17.70\rightarrow 14.60^{PS}$ & $20.70\rightarrow 17.20^{Iso.R}$ & $16.70\rightarrow 15.70^{Iso.R}$ & $18.50\rightarrow 16.50^{Iso.R}$ & $18.00\rightarrow 16.90^{Iso.R}$ \\
DE\_bald  & $15.60\rightarrow 12.40^{Iso.R}$ & $38.10\rightarrow 14.70^{PS/Iso.R}$ & $38.10\rightarrow 14.70^{PS/Iso.R}$ & $29.90\rightarrow 16.90^{Iso.R}$ & $27.50\rightarrow 15.00^{Iso.R}$ & $29.10\rightarrow 16.00^{Iso.R}$ & $33.20\rightarrow 15.80^{PS/Iso.R}$ \\
\bottomrule
\end{tabular}
}
\end{table*}

\subsection{Effectiveness of Calibration Metrics in Detecting Prediction Success}

Tables~\ref{vd_mece}--\ref{bs_dp} and Figures~\ref{dp2}--\ref{dp4} and \ref{vp2}--\ref{vp4} summarize how ECE, MECE, BS, and AUARC evaluate prediction reliability after uncertainty estimation and calibration. These metrics do \emph{not} provide interchangeable evidence in deep code models, however, they prioritize different properties of reliability, can rank the same model--metric pair differently, and can therefore support conflicting deployment decisions. In SE tasks, metric choice directly influences which predictions are trusted, deferred, or subjected to further review. 
\textbf{Metric disagreement is consequential in SE because it changes both model assessment and trust decisions.} The four metrics reward different properties of a score. ECE measures average score--frequency alignment, MECE emphasizes the largest local mismatch, BS combines probability error with class separation, and AUARC evaluates ranking quality for acceptance--rejection decisions. As a result, the same uncertainty signal can appear strong under one metric and weak under another. This effect is especially important in code tasks, where reliability is used operationally rather than descriptively. Practitioners do not only need a score that looks calibrated on average, but they need a score that helps identify which predictions remain safe to retain. Therefore, disagreement among metrics is itself a practical risk.


\textbf{ECE and MECE can misrepresent reliability in code models because both depend on how scores are distributed across bins.}
ECE can drop sharply without improving selective prediction. For example, in vulnerability detection, CodeBERT's entropy ECE decreases from \textbf{12.931\%} to about \textbf{2.194\%} after Iso.R, while AUARC stays near \textbf{0.754}. In defect prediction, MI for CodeBERT drops from \textbf{72.13\%} to \textbf{0.10\%} after Iso.R, yet its AUARC remains well below the strongest uncertainty signals. MECE is even less stable because it is determined by the worst populated bin and can be dominated by sparse score regions. Thus, in code models with skewed confidence distributions, both ECE and MECE can change substantially even when the set of risky accepted predictions remains largely the same. This makes them useful as diagnostics of score validity, but insufficient for deciding whether a model is safer to use.

\textbf{Bucket collapse is a mechanism by which calibration can improve aggregate metrics while weakening threshold-based decisions.}
Figure~\ref{bucket} illustrates this effect for Iso.R. Unlike TS and PS, Iso.R learns a non-strict monotone step function, so multiple distinct uncertainty scores can be mapped to the same calibrated value. This can improve bin-level calibration because each bucket is assigned an empirical correctness rate, but it also reduces score resolution. Predictions that were separable before calibration can become indistinguishable after calibration. This does not reverse the ordering across different buckets, but it removes the ordering within each collapsed bucket. As a result, selective prediction can become unstable when an accept--defer threshold falls inside a large bucket: the retained set may then depend on arbitrary tie-breaking rather than meaningful uncertainty differences. Thus, bucket collapse is not merely a visual effect; it indicates reduced resolution in the calibrated score space. While Iso.R does not reverse the ordering across buckets, it can eliminate within-bucket distinctions that selective prediction depends on, especially when an operating threshold intersects a large bucket.

\textbf{BS and AUARC answer different decision questions and neither is sufficient alone.}
BS can improve through global rescaling even when the score does not better separate correct from incorrect predictions. In vulnerability detection, calibrated BS values often concentrate near \textbf{0.215--0.236} across many signals, even though AUARC remains essentially unchanged. In defect prediction, MI-based BS improves substantially after calibration, but AUARC still trails the strongest confidence-based measures. AUARC, in contrast, is appropriate for evaluating selective prediction, yet it is largely insensitive to whether probabilities are numerically valid. Hence, BS may overstate reliability gains for deployment, while AUARC may understate whether the probabilities themselves are interpretable. In SE, these are distinct requirements where AUARC governs \emph{decision quality}, the BS governs \emph{probability meaning}.


\textbf{Code models require joint metric interpretation because reliability failures are concentrated in the retained region.} Figure~\ref{arc} provides an operating-point view of accepted-set risk. For a rejection rate \(r\), the retained coverage is \(1-r\), and the accepted-set error rate is \(1-\mathrm{Acc}(r)\), where \(\mathrm{Acc}(r)\) is the accuracy of the retained predictions. Thus, the curve does not only summarize ranking quality; it also shows the residual risk under different accept--defer policies. The key observation is that calibration changes the numerical scale of scores but does not substantially change the attainable risk--coverage tradeoff. In other words, after calibration, a practitioner may obtain better probability alignment while the same risky predictions remain in the accepted set at comparable coverage levels. The curves also become jagged at high rejection because decisions depend on a small retained subset, where minor ranking errors or score ties can determine which predictions survive thresholding.

\textbf{Takeaway.}
ECE, MECE, BS, and AUARC should not be interpreted as competing substitutes, but as complementary views of different reliability properties. Their disagreement is particularly important in SE because it affects three levels: \textbf{model comparison} depicts which code model appears most reliable, \textbf{decision support} depicts which predictions are trusted or deferred, and \textbf{metric interpretation} depicts whether reliability refers to probability validity or selective utility. For practitioners, the implication is that \textbf{no single metric should be used to justify deployment claims}. ECE and MECE should be treated as calibration diagnostics, AUARC should be used for trust/defer decisions, and conclusions about reliability should be drawn only from their joint behavior.

\begin{figure}[t]
    \centering

    \begin{subfigure}[t]{0.48\columnwidth}
        \centering
        \includegraphics[width=\linewidth]{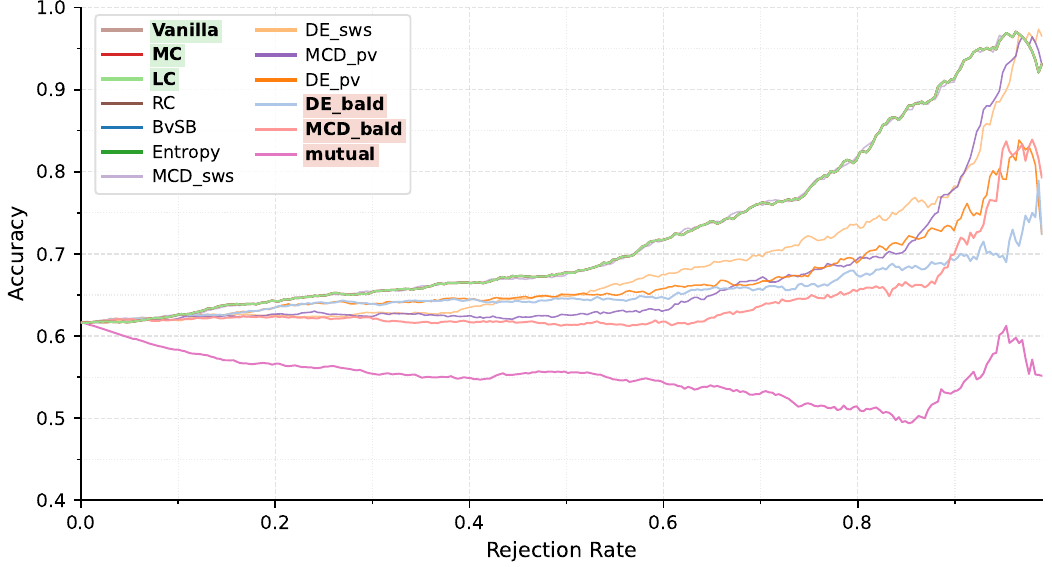}
        \caption{AUARC curve on uncalibrated model}
        \label{arc1}
    \end{subfigure}\hfill
    \begin{subfigure}[t]{0.48\columnwidth}
        \centering
        \includegraphics[width=\linewidth]{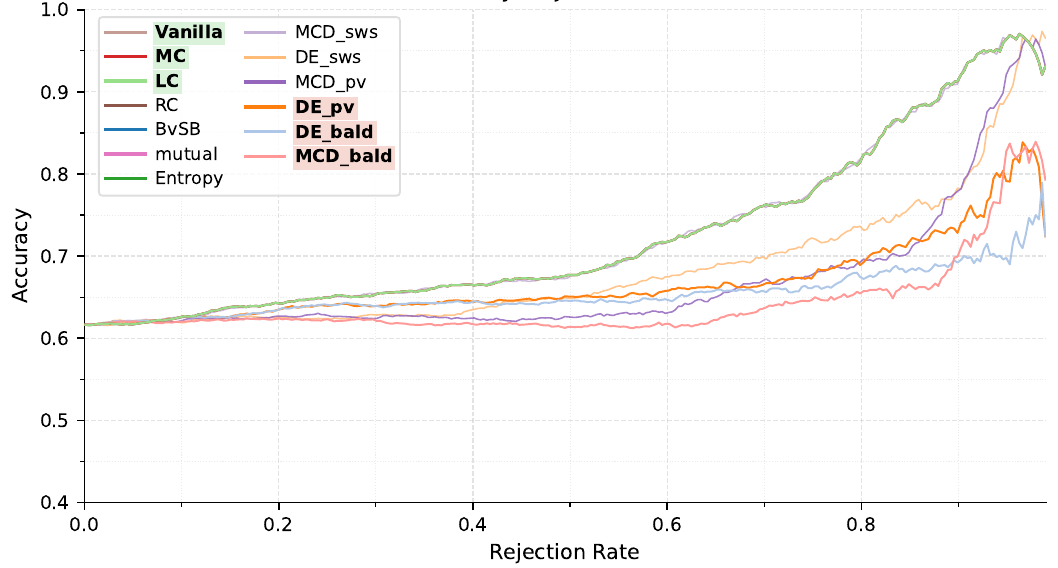}
        \caption{AUARC curve after applying PS}
        \label{arc2}
    \end{subfigure}\hfill

    \caption{Accuracy-rejection curve for vulnerability detection task on CodeLlama model before and after applying PS.}
    \label{arc}
\end{figure}

\subsection{Summary of Findings}

Across the three RQs, our results show that uncertainty reliability in deep code models cannot be characterized by a single metric. First, uncertainty scores can provide useful rankings for rejection while remaining poorly calibrated as probabilities. Second, post-hoc calibration can substantially improve score validity, as measured by ECE, MECE, or BS, without materially changing ranking-based selective-prediction behavior. Third, calibration metrics and selective-prediction metrics can disagree because they measure different properties of the same score: probability alignment, worst-case local mismatch, squared probability error, and accept--defer ranking quality.

Taken together, these findings indicate that calibration improvement should not be interpreted as a direct proxy for deployment reliability. A score may become numerically better aligned with empirical correctness while still failing to better separate correct from incorrect predictions. This distinction motivates the need to evaluate uncertainty in code models using multiple complementary views rather than relying on any single calibration or rejection metric.

\section{Operationalizing Reliability in SE Workflows}

Our findings suggest that uncertainty in code models should be operationalized as a mechanism for accept--defer decisions, rather than interpreted as a generic confidence score. In SE workflows, the relevant question is not only whether a model is calibrated on average, but whether its scores can support reliable decisions about which predictions may be trusted, deferred, or routed to additional analysis. This requires evaluating uncertainty under the intended operating regime.

\noindent\textbf{Define accepted-set risk.}
Practitioners should first specify the risk associated with predictions that the system will accept. This risk is task-specific. In vulnerability detection, the most harmful accepted error is a vulnerable function predicted as non-vulnerable; in defect prediction, different defect classes may carry different costs. Therefore, uncertainty should be evaluated using accepted-set risk, such as the error rate among retained predictions, class-conditional accepted error, or false-safe rate for security tasks. A model should not be considered reliable solely because its global ECE or Brier score improves.

\noindent\textbf{Select thresholds under an explicit operating target.}
Acceptance thresholds should be selected on validation data and evaluated on held-out test data. Given a target risk level, practitioners should choose the threshold that maximizes coverage while keeping the validation accepted-set risk below the desired limit. The selected threshold should then be applied to the test set without further tuning. This procedure makes the reliability claim operational: the model is not simply better calibrated, but supports a specific accept--defer policy at a measurable coverage and risk level.

\noindent\textbf{Separate probability validity from selective-prediction quality.}
Calibration and selective prediction should be used for different purposes. Calibration improves the interpretation of scores as correctness probabilities, while rejection curves measure whether scores rank reliable predictions ahead of unreliable ones. Because monotone calibrators preserve ranking, they can improve ECE or Brier score without improving AUARC or changing the composition of the retained set. Thus, a deployment claim should require both probability validity and selective-prediction quality.

\noindent\textbf{Inspect the retained region directly.}
Practitioners should examine the subset of predictions that would actually be accepted under the chosen threshold. Flexible calibrators such as isotonic regression may reduce calibration error while collapsing many examples into the same score bucket. Such bucket collapse can make threshold-based routing unstable, because many examples become indistinguishable to the decision policy. We therefore recommend reporting retained-region diagnostics, including accepted-set accuracy, high-confidence error rate, and the fraction of examples assigned to large calibrated buckets.

\noindent\textbf{Implication for SE research and practice.}
Uncertainty evaluation in SE should report three complementary dimensions: calibration quality, selective-prediction quality, and retained-region risk. Calibration metrics such as ECE and Brier score describe probability validity; rejection curves describe whether uncertainty supports accept--defer ranking; retained-region diagnostics describe the safety of the predictions that would actually be automated. Together, these measures turn uncertainty from a standalone model score into a software reliability policy for trust, deferral, and review decisions.

\section{Limitations}

Our evaluation is limited to two code classification tasks and seven code models. While vulnerability detection and defect prediction are important reliability-sensitive SE settings, the results may not fully generalize to other tasks, larger model families, or different deployment environments. We also do not study code generation, where uncertainty must account for token-level behavior, sequence-level correctness, decoding dynamics, and error propagation. Finally, our study focuses on post-hoc calibration and selective prediction, rather than training-time uncertainty objectives or deployment factors such as human review effort, latency, and tool-assisted follow-up analysis. These dimensions remain important directions for future work.

\section{Threats to Validity}

This section discusses potential threats to validity in our study, categorizing them into internal validity, which pertains to experimental design and methodology, and external validity, which concerns the generalizability of our findings.


\textbf{Internal Validity: }
A potential limitation of our study is the selection of uncertainty metrics and calibration methods. We evaluate 13 techniques~\cite{hendrycks2018baseline,steinhardt2016unsupervised,shannon1948mathematical,monarch2021human,gal2016dropout,lakshminarayanan2017simple,CodeImproveicse}, alongside post-hoc methods such as TS, PS, and Iso.R. While our comparisons systematically assess effectiveness, certain techniques may interact differently with specific model architectures, potentially biasing results. Further validation with additional calibration methods is needed to ensure the robustness of our conclusions.
Architectural and pretraining differences can affect how uncertainty is expressed and how well scores calibrate. While we evaluate a diverse set of models up to 7B parameters, broader validation on larger models and additional datasets is needed to confirm generality.

A key challenge in uncertainty calibration is bin imbalance during error computation. Fixed-width binning assumes uniformly distributed confidence scores, which rarely holds for real-world models with skewed uncertainties. We do not evaluate adaptive binning methods (e.g., dynamic or density-based)~\cite{nixon2019measuring,balanya4379258adaptive,wang2023calibration}, although they may better characterize calibration in code models.

\textbf{External Validity:} 
The generalizability of our findings focus on experiments on vulnerability detection and defect prediction~\cite{zhou2019devign,phan2017conv}. While these address practical SE challenges, they may not fully extend to tasks like code summarization, bug localization, or automated program repair.

The datasets we use are drawn from specific open-source projects and programming platforms, which may not fully represent all real-world software development scenarios. If these datasets contain inherent biases (e.g., an overrepresentation of certain coding patterns or vulnerabilities), our models may learn spurious correlations, affecting uncertainty estimates. Evaluating our findings on more diverse datasets, including proprietary and industry-scale codebases, would provide stronger external validation.

Real-world deployment considerations pose a challenge. While our study evaluates uncertainty calibration in an offline experimental setting, actual deployment scenarios introduce additional factors such as concept or data distribution shifts, and adversarial code modifications. Since models in production environments continually encounter previously unseen code patterns, their uncertainty estimates degrade over time. Future work should assess how calibration techniques adapt to evolving software landscapes to maintain reliability in dynamic development environments.


Our study takes an initial step in identifying key gaps in existing calibration techniques and outlining research directions to improve decision-making in code analysis. We acknowledge these limitations as future work, 
recognizing the need for more adaptive calibration techniques and evaluation strategies that go beyond statistical calibration to enhance practical uncertainty separability. 

\section{Related Work}\label{sec::related}

\textbf{Pre-trained Code Models.}
Pre-trained code models, primarily based on the Transformer architecture, fall into three categories: encoder-only, decoder-only, and encoder-decoder models. Encoder-only models (e.g., CodeBERT , GraphCodeBERT ) focus on code understanding~\cite{fengCodeBERTPreTrainedModel2020,guo2021graphcodebert,Liu2019RoBERTa}, while decoder-only models (e.g., CodeGPT, GPT-C ) are designed for generative tasks like code completion~\cite{achiam2023gpt,tian2023chatgpt}. Encoder-decoder models (e.g., PLBART, CodeT5) integrate both components for comprehensive programming language modeling~\cite{wang2021codet5,ahmad2021unified}. Pre-trained on large-scale code corpora, these models provide transferable representations that can be fine-tuned for downstream tasks such as defect prediction and vulnerability detection.

\textbf{Model Calibration.}
Model calibration has been extensively studied in NLP~\cite{chen2022close,nguyen2015posterior} and CV~\cite{guo2017calibration,wang2023calibration,balanya4379258adaptive,nixon2019measuring,laves1909well,krishnan2020improving} but remains underexplored for code models. Also, research focused on improving calibration through temperature scaling and platt scaling in for code models~\cite{spiess2024calibration,zhou2024calibration}. Later studies evaluated LLM calibration across NLP tasks, revealing improved calibration with model scaling and pretraining, though some findings remain disputed. Some works investigated uncertainty estimation in code models ~\cite{vasconcelos2022generation,johnson2023ru}, while others explored LLMs' self-assessment of correctness~\cite{kadavath2022language,key2022toward}. Techniques such as Bayesian approaches~\cite{minderer2021revisiting}, mixup~\cite{park2022calibration}, and perturbation-based methods~\cite{huang2023look} have been explored for calibration improvements. However, calibration in software engineering tasks remains largely uncharted. To our knowledge, this is the first systematic study of uncertainty calibration for selective prediction in code classification. 
Our work differs by jointly evaluating calibration quality, selective-prediction quality, and retained-region risk for accept--defer decisions in code classification tasks.

\section{Conclusion}\label{sec::conclusion}

In this paper, we examined the reliability of uncertainty estimates and post-hoc calibration for deep code classification models in vulnerability detection and defect prediction. Across a diverse set of modern code models, we found that standard uncertainty scores and widely used calibration methods can improve aggregate calibration measures yet still leave a critical failure mode unresolved: incorrect predictions that retain high confidence, which undermines selective prediction in deployment settings. These results show that numeric probability alignment alone is not sufficient for trustworthy accept–defer decisions, and that evaluation must account for both calibration quality and abstention behavior. 

\clearpage

\bibliographystyle{IEEEtran}
\bibliography{sample-base}

\end{document}